\documentclass[review,number,sort&compress]{elsarticle}
\usepackage{lineno}

\usepackage{epsfig}

\usepackage{amssymb}
\usepackage{float} 
\journal{NIM A} 

\begin{document}

\begin{frontmatter}



\title{An elastic lidar system for the H.E.S.S. Experiment}


\author{J. Bregeon, M. Compin, S. Rivoire, M. Sanguillon, G. Vasileiadis \footnote{Corresponding author:   george.vasileiadis@lupm.in2p3.fr}}

\address {LUPM, IN2P3/CNRS and Un.Montpellier II}
\begin{abstract}
 The H.E.S.S. experiment in Namibia, Africa, is a high energy gamma ray telescope sensitive in the energy range from $\sim$100\,Gev to a few tens of TeV, via the use of the atmospheric Cherenkov technique. To minimize the systematic errors on the derived fluxes of the measured sources, one has to calculate the impact of the atmospheric properties, in particular the extinction parameter of the Cherenkov light ($\sim$300-650\,nm) exploited to observe and reconstruct atmospheric particle showers initiated by gamma-ray photons. A lidar can provide this kind of information for some given wavelengths within this range.  In this paper we report on the hardware components, operation and data acquisition of such a system installed at the H.E.S.S. site.
\end{abstract}


\end{frontmatter}


\section{Introduction}
The H.E.S.S. (High Energy Stereoscopic System) experiment consists of 5 imaging Cherenkov telescopes situated in the Namibia Khomas Highland desert (1800\,m asl). Its main objective is the study of galactic and extragalactic sources in the energy range of $\sim$100\,GeV to a few tens of TeV coupled to a substantial flux sensitivity (1$\%$ Crab units). The combination of the 5 telescopes data analysis provides good background rejection and angular resolution. The detection technique used by H.E.S.S., namely the measurement of Cherenkov light emitted from secondary particles created 5 to 10 \,km above ground level, demonstrates by itself the importance of knowing any variation of the atmospheric quality \cite{H.E.S.S.}. 

Precise flux and energy spectra calculations for the observed sources could suffer if the level of  extinction of cherenkov-generated photons varies due to aerosols or thin particles present in the atmosphere during the gamma shower development. Similar experiments use by default the uniform rate of background cosmic-ray initiated showers as a measurement of atmospheric clarity\cite{CTC}. To obtain more precise estimations additional instruments such as radiometers, transmissometers or lidars can be implemented. Operating a lidar during the data-taking phase permits us to model the atmospheric transmission above the site, which in sequence could be used to simulate different background conditions by varying the aerosol density, much like the real background data.  Once this is achieved, these transmission tables are used to produce corrected tables for energy and effective area by means of gamma-ray simulations.
\begin{figure}
   \centering
   \includegraphics[width=8.0cm]{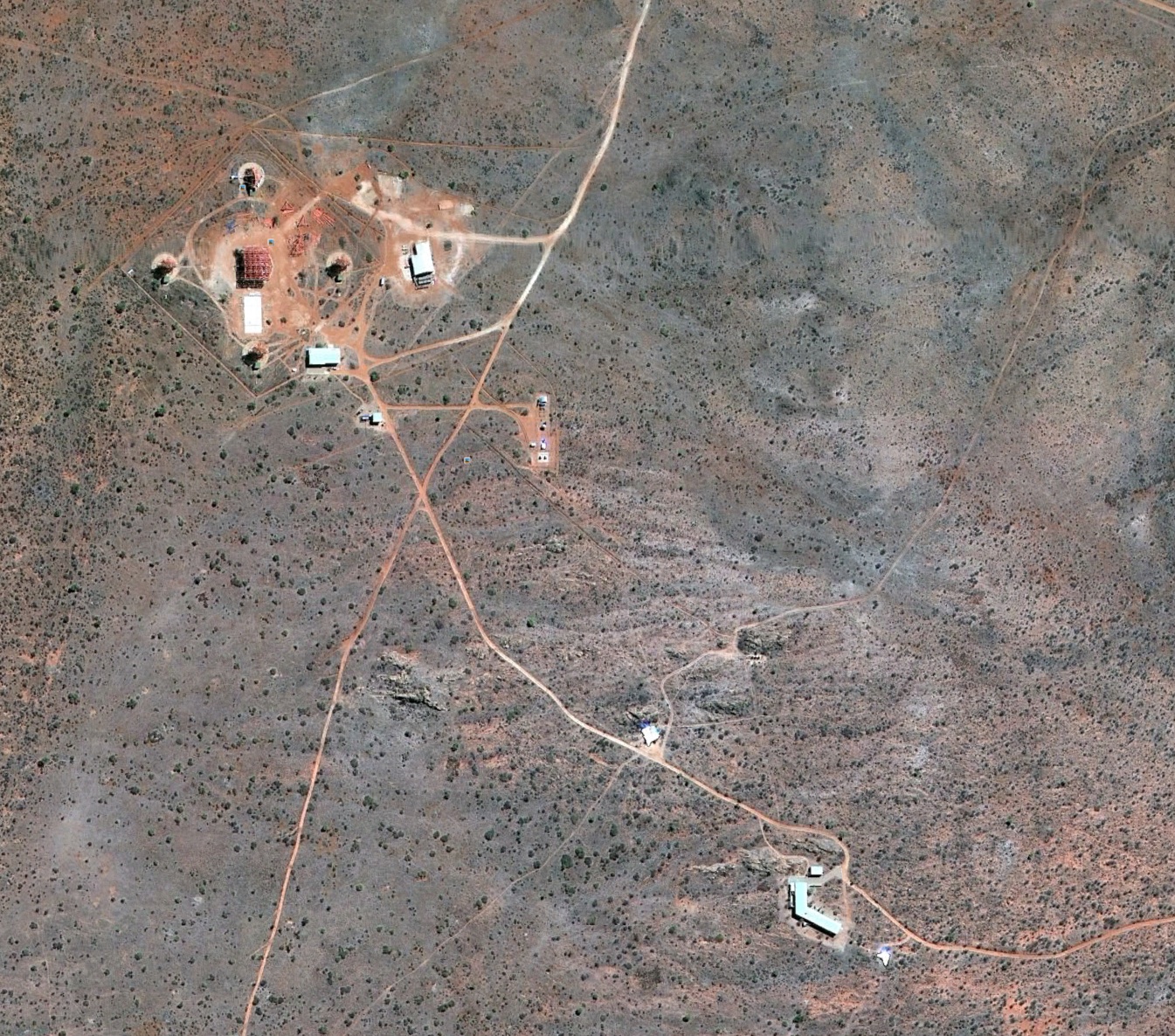}
      \caption{Aerial view of the H.E.S.S. site, the lidar hut is visible at the lower right part of the photo. The distance between the telescopes and the lidar been 850m to avoid any interference due to the laser shots.}
         \label{fig1}
  \end{figure}

This paper describes the design, operation and performance of a lidar system installed at the H.E.S.S. experiment site, fig.\,1. It is organized as follows. We give a summary of the relevant atmospheric parameters in section 2. Section 3 describes the current lidar hardware. In section 4, the signal detection and processing is presented. First results of the analysis of the atmospheric parameters are summarized in section 5. A summary concludes the paper. 
\begin{figure}[h]
   \centering
   \includegraphics[width=5.cm]{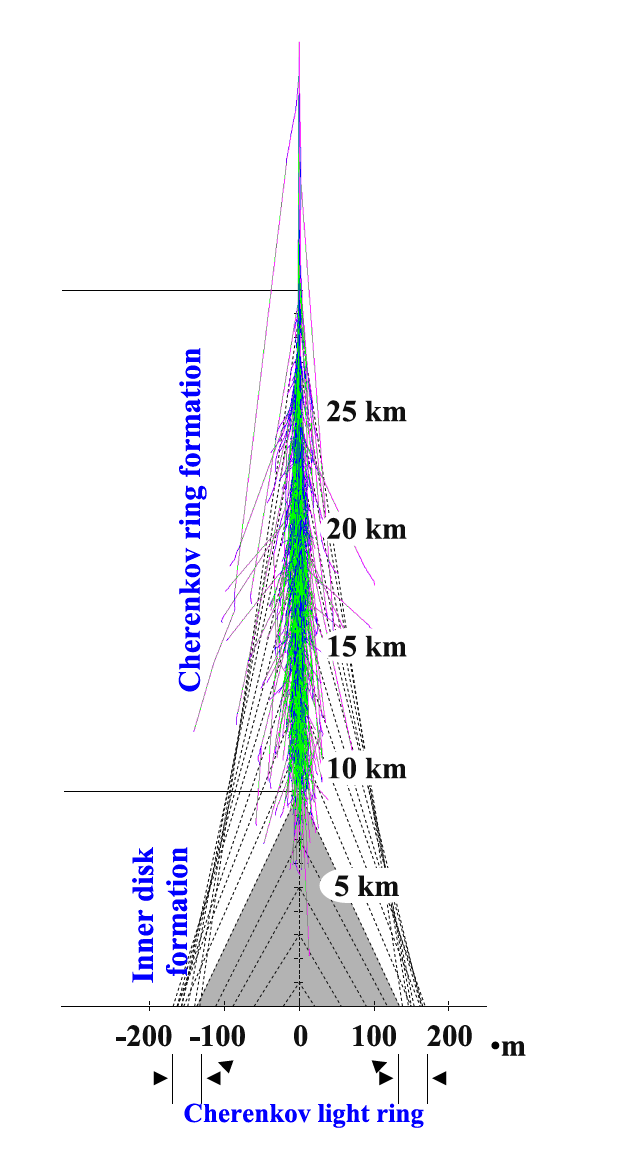}
      \caption{Characteristics of a Cherenkov shower development of an incoming $\gamma$ ray particle in the atmosphere.}
         \label{fig2}
   \end{figure}
\section{Atmospheric Monitoring}
In order to study very high energy cosmic rays, the atmosphere can be used as the Cherenkov radiator, while the radiation is imaged onto single photon sensitive detector planes of telescopes with large parabolic mirrors. A  $\gamma$ ray interacting in the atmosphere will produce an air shower of secondary particles at an elevation of 10-15\,km. The shower particles  produce Cherenkov light and the resulting light cone will cover a disk of a radius of 100-120\,m at the ground level with an intensity of 10-100 photons m$^{-2}$,  fig.\,2. The telescopes looking up the night sky will be able to detect the Cherenkov light and measure the intensity, orientation and shape of the air shower, which is related to the primary energy and direction of the  $\gamma$ ray.

The loss of Cherenkov light from a shower (as viewed by the H.E.S.S. telescopes) is mainly due to molecular and Mie scattering. The presence of aerosols can affect the data two-fold. If close to ground level the recorded photon yield can be lower affecting the telescope trigger thus affecting the reconstructed shower energy. If close to the shower maximum, at around 10km height ,  the shape and brightness of the camera images would be affected. The introduction of lidar measurements could permit the calculation of the extinction parameter of the atmosphere $\alpha$ thus reducing the systematic error on the final energy spectrum calculation.

The back-scattered signal emitted by a lidar, after being back-scattered by the atmosphere and collected by a telescope, is measured by a photomultiplier tube (PMT). The light pulse height from a distance R is given by the lidar equation
\begin{equation}
P(R)=K\frac{\beta(R)}{R^{2}}e^{-2\tau(R)}
\end{equation}
where $\beta(R)$ is the back-scattering coefficient and $\tau(R)$ is the integral of the extinction coefficient $\alpha(r)$ along the path. Both these quantities are sums of the aerosol and molecular contributions: $\alpha(R)=\alpha_{mol}(R)+\alpha_{aer}(R), \beta(R)=\beta_{mol}(R)+\beta_{aer}(R)$. The molecular part can be evaluated either using standard models, knowing the pressure and temperature vs. height from atmospheric measurements or radiosonde data from nearby meteorological sites. On the other hand the extraction of the aerosol coefficients, based on the lidar equation, demands algorithms based on either the Klett \cite{klett} or Fernald \cite{fern} methods. Since both methods need a priori assumptions about the molecular vs. aerosol parameter, the calculated value of $\alpha$ comes with a relatively important systematic error. For the purposes of this paper we have used the Fernald approach to present a preliminary analysis of our data while to evaluate the molecular contribution we used also radiosonde data available for our site.
\begin{figure}[h]
   \centering
   \includegraphics[width=7cm]{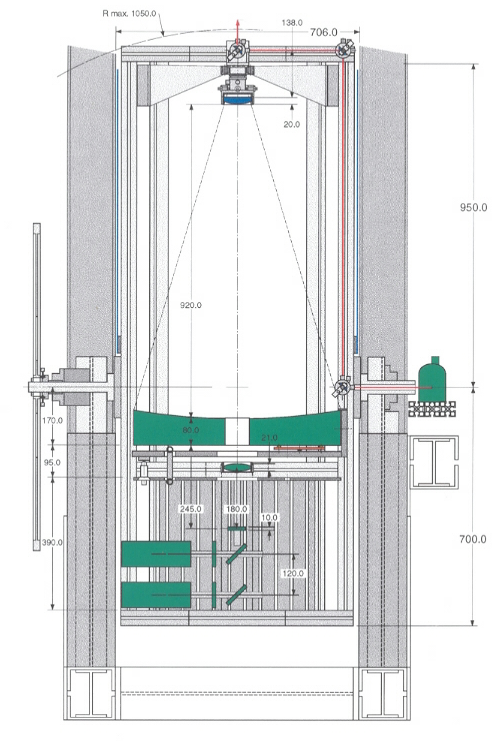}
      \caption{Schematic diagram of the H.E.S.S. lidar System. The Quantel laser is mounted on the right-hand side of the telescope, while a pair  of photomultipliers are situated behind the primary mirror and at a distance of few cm further away than the optimal focal point of thje telescope  to achieve a more uniform illumination of the photomultplier's cathode.}
         \label{fig3}
   \end{figure}
\section{Lidar Hardware and Data Acquisition}
The lidar telescope was conceived in 1997 and construction was completed in 1999. Until 2001 it was installed at the Themis site (Pyrenees,France) for the purpose of the Celeste experiment \cite{celeste}. A mechanical aspect of the telescope structure and light collection is sown in figure 3. Due to poor weather conditions on site it was used only for short intervals which led to frequent mechanical failures. 

During summer 2004 it was transferred and installed at the H.E.S.S. site, in Namibia (Africa). It was installed in a dedicated hut, situated at a distance of 850\,m from the H.E.S.S. facilities, minimizing the possibility that the cameras of the H.E.S.S. telescopes sees the lidar laser beam. After a complete rebuild, involving replacement of obsolete parts and software upgrades, it started routine data taking operations during the summer of 2006. 
\subsection{Mirror and Mount}
For the collection of the backscattered light, the lidar uses a Cassegrain type telescope, fig. \,3. The primary parabolic mirror is of $\Phi=60$\,cm diameter with a focal length of $\phi=102 cm$, while the secondary has a diameter of 8\,cm and $\phi=10$\,cm. The mirrors were produced by Compact using BK7 glass coated with aluminum and a reflectivity of $80\%$ in the range 300-600\,nm. The average spot size in the focus is 1.5\,mm FWHM. It is mounted in a fully steerable alt-azimuth frame equipped with DC servomotors with a maximum speed of $5^{o}$/s. The absolute pointing direction is close to $0.7^{o}$ accuracy. A three point mounting system allows for alignment and collinearity of the mirror and laser beam.

The whole apparatus is installed in a 5$\times$5\,m hut equipped with a motorized roof, protecting it from rain and harsh conditions when the lidar is not in operation. All motors can be controlled either locally or remotely via Ethernet connections.
\subsection{Laser}
The choice of the laser for our purpose is dictated by the following requirements: the wavelength of the laser has to match as much as possible the observed Cherenkov photons energy spectrum (300-650\,nm). It follows that a double wavelength laser is needed, giving us two reference points to compare with this spectrum. Repetition rate should be high enough to reduce collection time, while the laser power should be adjustable to avoid interference with the H.E.S.S. telescope optical systems.
To meet these requirements a Quantel Brilliant 20 Nd:YAG laser was used. It is equipped with two cavities generating the 2nd and 3rd harmonics at 532 and 355\,nm. The repetition rate is 10\,Hz while the per-pulse energy is 180\,mJ and 65\,mJ respectively. The laser is mounted aside the telescope structure. It is guided in a bi-axial configuration at a distance of 43\,cm from the optical axis of the telescope.

\subsection{Signal Detection, Trigger and Digitization}
A pair of Photonis XP2012B photomultipliers are used for backscatter light detection. The return signal is split by means of a dichroic filter, mounted at the focal point, separating the 532\,nm from the 355\,nm component. We use the photon counting method as a measurement which results in a saturation of the signal for the first few hundred meters. The photomultiplier gain reaches the value of $3\times10^{6}$ at 1450\,V. The whole 51\,mm-diameter photocathode window is exposed to maximize light detection.

The output signals are fed through a pair of 1.5\,m long cables to a two-channel 12\,bit Compuscope Octopus CAGE 8265 Digitization board that runs at 65\,Mhz. This gives us an altitude resolution of 2.5\,m. A series of PCI boards command the servomotors and hydraulic brakes. Finally a Labview interface supervises the whole sequence of steering and data-taking of the lidar system on a Windows XP based industrial CPU.

The laser system is being driven by means of a client-server protocol. The server is running on the H.E.S.S. DAQ system while the dedicated lidar CPU runs the client part. Upon reception of an acquisition demand, accompanied by several parameters such as pointing direction, laser power and run duration, the client executes the required configuration , points the telescope and triggers the laser system up to the nominal laser repetition rate. The DAQ is triggered by the laser Q-switch synchronization pulse generated at every successful laser shot.

\subsection{Operation Mode and Maintenance}
The lidar system can be operated in two ways. Either in stand-alone mode or in slave mode where it is being driven by the H.E.S.S. central data acquisition system. The idea was that during the first months of operations we used the standalone mode for a complete debugging and data analysis optimization while later we switched to slave mode for normal data taking. During normal H.E.S.S. operation the dedicated lidar runs are executed within the time interval that separates the consecutive physics runs, a 180 sec time interval. A 1200 laser shots profile is executed every time and stored in the H.E.S.S. database in form of a ROOT file \cite{root}. 

During the commissioning phase of the lidar, it was pointed out by the H.E.S.S. experiment that the option to steer the laser beam of the lidar could cause some issues with the operation of the Cherenkov telescopes. The danger being that the laser beam crosses the field of view of the telescopes while in operation, resulting in damage to the telescope's camera readout system. It was decided then  to point the lidar telescope and laser beam in a fixed direction ($75^{o}$ azimuth and $25^{o}$ WE direction) to avoid any overlap between the two experiments.

Maintenance was centered around the re-alignement of the optical system. This operation was repeated once a year to assure optimal operation of the lidar. At the same occasion the laser flash lamp was also being replaced even if the typical lifetime ($>50$ millions shots) was not reached. The entire de-ionized water cooling system was also being serviced and filters were replaced on a yearly basis as well. Since during winter time the temperature could drop below zero during the night, the laser temperature head was kept between 5 and 30 degrees by means of a temperature regulated air flow system.
\begin{figure}
   \centering
   \includegraphics[width=11cm]{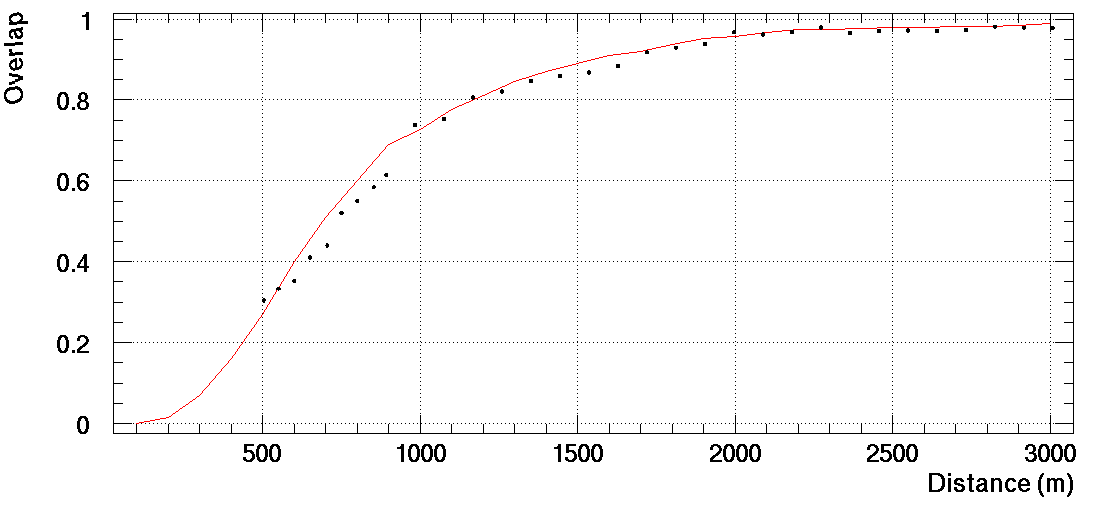}
      \caption{Backscatter signal obtained from a distance up to 3000\,m using horizontal laser acquisitions. Solid line correspond to simulated  performance using the OSLO optical design software for our setup \cite{oslo}.} 
         \label{fig4}
   \end{figure}

\subsection{Signal Treatment}
The recorded signal is described by the so-called lidar equation :
\begin{equation}
P(r)=P_{0}\frac{ct_{0}}{2}\beta(r)\frac{A}{r^{2}}e^{-2\tau(r)}=P_{0}\frac{ct_{0}}{2}\beta(r)\frac{A}{r^{2}}e^{-2\int^{r}_{0}a(r')dr'}
\end{equation}
where $P(r)$ is the signal received at time $t$ from photons scattered at a distance $r$ from the lidar, $P_{0}$ is the transmitted laser power, $t_{0}$ is the laser pulse duration, $\beta(r)$ is the backscattering coefficient, $\tau(r)$ is the optical depth, $\alpha(r)$ is the extinction coefficient, and $A$ is the effective area of the detector. The latter is proportional to the overlap function of the telescope.

Using the experimental method described in \cite{overp}, we calculated the overlap function using horizontal scans. The results (shown in fig.\,4) are compatible with an analytical calculation regarding the minimum useful observation altitude, calculated at around 2000\,m. For analysis purposes we used an initial altitude of $r_{0}$=800\,m above site due to signal saturation due to stray light entering the telescope tube, still we have applied the overlap  correction factor shown above for the rest of the data points.

Data after background subtraction, calculated in the range of 20-25\,km, where we expect no laser signal, were smoothed using the typical Savinsky-Golay algorithm. Resolution was also reduced from the initial 2.3\,m to 92\,m using the gliding window technique. This was necessary to improve the statistics per altitude unit but also to be coherent with the expected Monte Carlo resolution used with the H.E.S.S. energy spectra reconstruction algorithms. On the latter the atmosphere is divided in layers of a given thickness (200\,m) and all effects due to aerosol or molecule presence are bein g studied using this minimum path length. 
 \begin{table}
\centering
\caption{Typical Lidar ratio values at 532\,nm and different aerosol types}
\begin{tabular}{lllll}
\\ \hline
Marine particles          & 20-35 sr &  &  &  \\
Sahara dust               & 50-80 sr &  &  &  \\
Urban particles           & 35-70 sr &  &  &  \\
Biomass burning particles &   70-100 sr  &  &  & 
\\ \hline
\end{tabular}
\end{table}

\section{Analysis method and data sets}
Backscattering and extinction are both caused by particles and molecules. The first contribution is related to aerosol and cloud presence, while the second is due to Raleigh scattering. It follows that we can write from equation (2) :
\begin{eqnarray*}
\beta(r)=\beta_{aer}(r)+\beta_{mol}(r) \\
\alpha(r)=\alpha_{aer}(r)+\alpha_{mol}(r)
\end{eqnarray*}
The molecular scattering part can be determined using pressure and temperature profiles from meteorological measurements or approximated from appropriate atmospheres, in our case desert type. So the only part that needs to be determined is the aerosol scattering part. It is common to introduce two ratios in this case, the particle extinction-to-backscatter ratio (lidar ratio) 
\begin{equation}
L_{aer}(r)=\frac{\alpha_{aer}(r)}{\beta_{aer}(r)}
\end{equation}
and in similar manners the molecular lidar ratio 
\begin{equation}
L_{mol}(r)=\frac{\alpha_{mol}(r)}{\beta_{mol}(r)}=\frac{8\pi}{3}sr
\end{equation}
Typical values for the lidar ratio $L_{aer}$ ( shown in Table 1) have been discussed in the literature \cite{profi}. 

Taking into account these considerations and after some integrations, equation 2 takes the following Bernoulli type form :
\begin{eqnarray*}
\frac{d ln(S(r)L_{aer}(r) exp\left \{ -2 \int_0^r (L_{aer}(r')-L_{mol}) \beta(r')_{aer} dr' \right \})}{dr}  \\
=\frac{1}{Y(r)}\frac{dY(r)}{dr}-2Y(r)
\end{eqnarray*}

where S(r) corresponds to the the range-corrected signal $S(r)=R^2P(r)$ and $Y(r)=L_{aer}(r) \left \{\beta(r)_{aer}(r)+\beta(r)_{mol}(r)\right \}$. To resolve this equation we follow the Fernal method, assuming a reference range where no aerosol contribution to the backscattered signal is expected and integrating from then on (backward integration).

For the last two and a half years we have operated our lidar in a regular base following the observation schedule of the H.E.S.S. experiment. A total of 2650 profiles, both at 355\,nm and 532\,nm, were accumulated covering a wide variety of weather conditions. Most of the datasets concern the Namimbian winter periods, June to October, where the H.E.S.S. experiment is most active and naturally only during night time. The results presented here are based on a subset of these data after omitting nights where weather conditions were bad enough that H.E.S.S. observations weren't possible, namely rain, heavy winds or complete overcast skys. On the other hand our lidar was used intensively when aerosol levels were elevated or thin clouds were present. A particular case where we have focused our efforts concerned the period where fires and other biomass burning activities are very active in the region. Such events are expected to influence the total trigger rate of the H.E.S.S. telescope so a dedicated campaign was conducted.
\begin{figure}
   \centering
   \includegraphics[width=12cm]{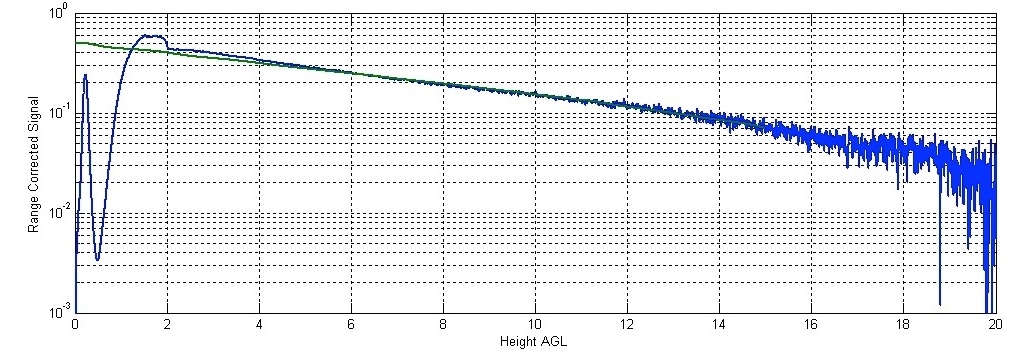}
      \caption{Normalized range corrected return signal (blue line). Green line correspons to the molecular corss section calculated profile.}
         \label{fig5}
  \end{figure}

\section{Results}
\begin{figure}[h]
   \centering
   \includegraphics[width=11cm]{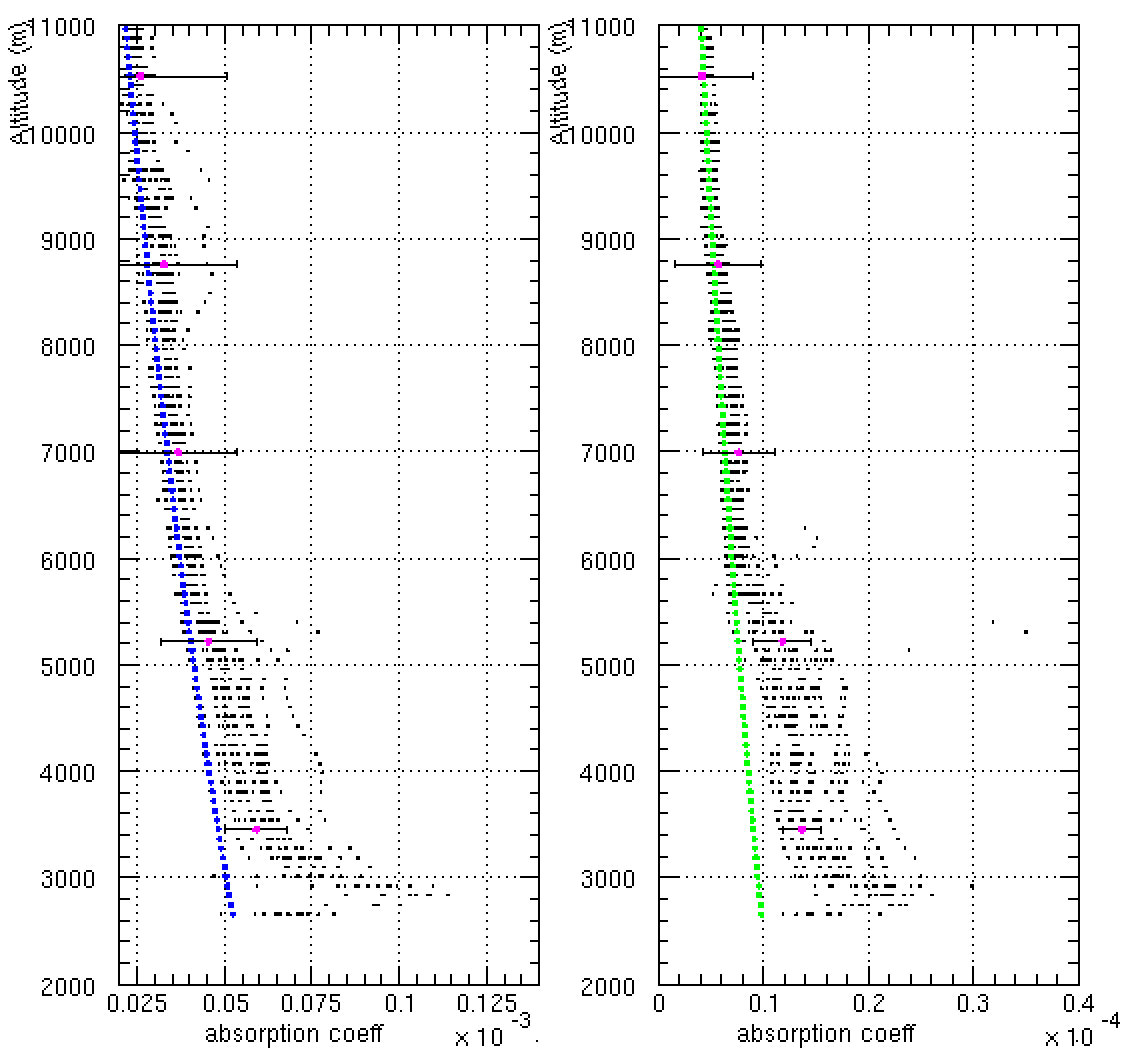}
      \caption{Typical extinction profiles at 355 (left) and 532 (right) nm obtained during summer 2013 on site. Solid lines corresponds to Rayleigh scattering after pressure and temperature from radio sonde profiles taken at a nearby station.}
         \label{fig6}
  \end{figure}

\subsection{Atmospheric transparency} 
The lidar range-corrected signals were first compared and normalized to the molecular scattering cross section profile computed from measured pressure and temperature on site. A suitable region where no aerosol or cloud presence was expected was chosen for this purpose, typically between 6 and 8\,km above site. A typical case is shown in fig.\,5. Following this, a reference range-bin has been set as the calibration point required for the Klett-based analysis. The altitudinal distribution of $\alpha_{obs}$ was then calculated. 

Examples of the extinction coefficient obtained from these measurements and for various conditions are shown in fig.\,6. We use data points up to 11\,km since above this altitude the return signal was too weak. The solid line on these plots represents the $\alpha_{Ray}$ calculated for pure Raleigh scattering. We used a lidar ratio of 70 for all our analyses, a value that corresponds to desert conditions, see table 1. Error bars indicate one standard deviation caused by statistical uncertainties and have been calculated from the law of error propagation by assuming a Poisson noise distribution on lidar signals. 

\begin{figure}
   \centering
   \includegraphics[width=12cm]{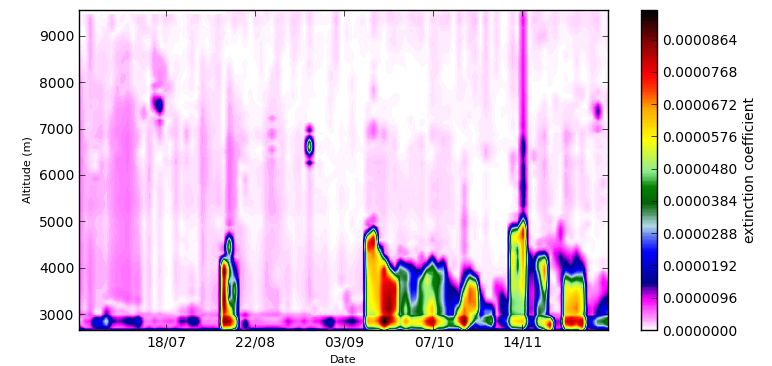}
      \caption{Timeline of the extinction coefficient measured during summer 2013. A biomass type burning effect is visible towards the end of this period. Extensive presence of dust particles is clearly  seen up to altitudes of 5km above sea level. }
         \label{fig7}
  \end{figure}

As it was mentioned before, an interesting effect during the winter period in Africa is the biomass burning effect. This is mainly due either to fires been initiated by farmers to enrich their fields or by accidental or weather-related reasons. High concentrations of smoke and dust produce relatively significant concentrations of aerosols at low altitudes. 

The timeline plot shown in fig.\,7 represents extinction coefficients calculated for the period of 6 June to 24 November 2013. Data were taken every second night, accumulating on an average 8-10 profiles per night, evenly distributed throughout the night period. Extinction was measured on both wavelengths 355\,nm and 532\,nm (only 532\,nm shown). A strong concentration of low altitude aerosols and dust, up to 3000\,m above the lidar horizon, has being observed between the 5th of September and 16th of November, a period that coincides with the above described local activities

\begin{figure}[h]
   \centering
   \includegraphics[height=8cm,width=11cm]{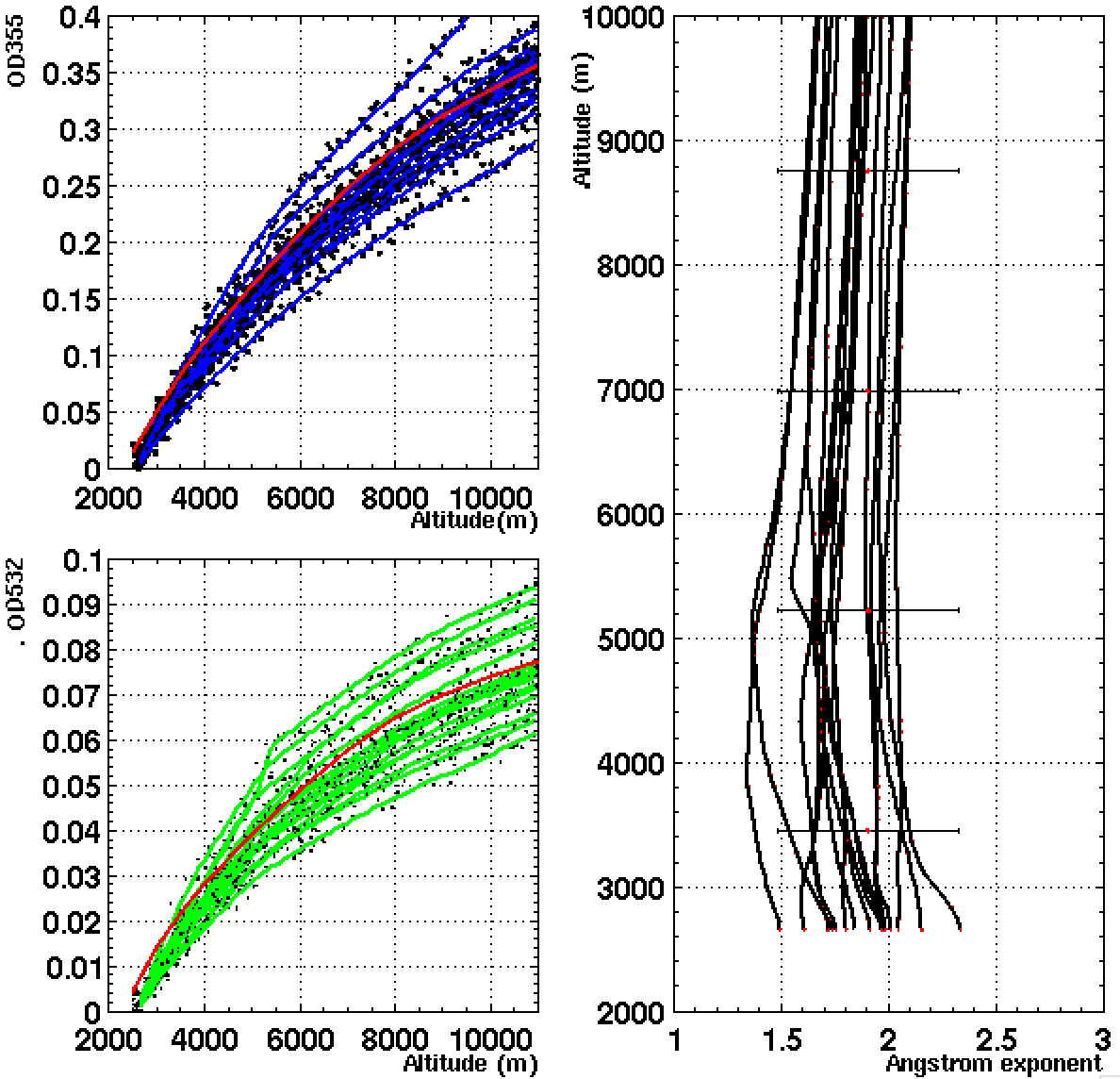}
      \caption{Variations of the optical depth based on datasets of winter 2013. Red line is based on MODTRAN5 typical desert aerosol profile on the right side the Ansgtrom exponent 532nm/355nm for the same period. Particle size corresponds to that due to biomass burning effect.}
         \label{fig8}
  \end{figure}

\subsection{Vertical Aerosol Optical Depth and Angstrom exponent}
In this section we evaluate the opacity of the atmosphere in terms of the Aerosol Optical Depth (AOD). The AOD for a given height r, $\tau_{AOD}$, is defined as the integral of the extinction coefficient $\alpha_{obs}$ between the Lidar ground level and r as follows :
\begin{equation}
\tau_{AOD}(r)=\int_0^r\alpha_{obs} (r')dr'
\end{equation}
The median of $\tau_{AOD}$(r) for the 355\,nm and 532\,nm line as a function of altitude is shown on fig.\,8. This covers the time period mentioned above. Superimposed, red line, is a typical profile expected using the  a MODTRAN5 \cite{mod5} simulation package. We used a mixture of desert climat with aerosol presence parametrization, while the atmospheric conditions corresponded to the mean vlaues ones registered on site for this specific period.  Overall accordance is satisfactory, while the variations of the measured AODs are expected since cloud and dust presence on site is know to vary substantially with a single night.

Using the above obtained AODs for these two wavelengths we were able to calculate the Angstrom exponent 532\,nm/355\,nm shown on fig.\,8, right side plot. Values vary between 1.3 and 2.2, corresponding to biomass burning types of particles, in accordance with the observed period and related activities in the region.

\section{Conclusions}
We've described the H.E.S.S. elastic lidar, and explain operation and maintenance matters. We've shown that the lidar works well, provides good quality data and that we're able to analyse them to show the presence of biomass burning aerosols on site.

Next steps are to fine tune the analysis, show the impact of aerosols on the instrument response, correct the instrument response using atmospheric transmission from lidar data in simulations, and show we can retrieve Cherenkov data for not perfectly clear nights, and demonstrate the validity of the procedure on one or more bright sources, see for example \cite{magic}.
All this we'll be published in a near future. Main issue of the elastic lidar is that you have to assume an aerosol type, what leads to quite large uncertainties, that's why we're working on a Raman lidar for CTA.


\end{document}